\documentstyle[12pt]{article}
\catcode`@=11     % make @ like a letter
     
\newdimen\jot \jot=3pt
\newskip\z@skip \z@skip=0pt plus0pt minus0pt
\newdimen\z@ \z@=0pt % can be used both for 0pt and 0
\dimendef\dimen@=0
     
\def\m@th{\mathsurround=\z@}
     
\def\ialign{\everycr{}\tabskip\z@skip\halign} % initialized \halign
     
\def\openup{\afterassignment\@penup\dimen@=}
\def\@penup{\advance\lineskip\dimen@
  \advance\baselineskip\dimen@
  \advance\lineskiplimit\dimen@}
     
\def\eqalign#1{\null\,\vcenter{\openup\jot\m@th
  \ialign{\strut\hfil$\displaystyle{##}$&$\displaystyle{{}##}$\hfil
      \crcr#1\crcr}}\,}
\def\meqalign#1{\null\,\vcenter{\openup\jot\m@th
  \ialign{\strut\hfil$\displaystyle{##}$&&$\displaystyle{{}##}$\hfil
      \crcr#1\crcr}}\,}
     
\catcode`@=12   % back to nonletter category
\newcommand{\be}{\begin{equation}}
\newcommand{\ee}{\end{equation}}
\newcommand{\bea}{\begin{eqnarray}}
\newcommand{\eea}{\end{eqnarray}}

\newfam\bofam
\font\tenbo=cmmib10   \textfont\bofam=\tenbo

\mathchardef\Omega="710A
\mathchardef\alpha="710B
\mathchardef\beta="710C
\mathchardef\gamma="710D
\mathchardef\delta="710E
\mathchardef\epsilon="710F
\mathchardef\rho="711A
\mathchardef\sigma="711B
\mathchardef\tau="711C
\mathchardef\upsilon="711D
\mathchardef\phi="711E
\mathchardef\chi="711F
\mathchardef\Gamma="7100
\mathchardef\Delta="7101
\mathchardef\Theta="7102
\mathchardef\Lambda="7103
\mathchardef\Xi="7104
\mathchardef\Pi="7105
\mathchardef\Sigma="7106
\mathchardef\Upsilon="7107
\mathchardef\Phi="7108
\mathchardef\Psi="7109
\mathchardef\zeta="7110
\mathchardef\eta="7111
\mathchardef\theta="7112
\mathchardef\iota="7113
\mathchardef\kappa="7114
\mathchardef\lambda="7115
\mathchardef\mu="7116
\mathchardef\nu="7117
\mathchardef\xi="7118
\mathchardef\pi="7119
\mathchardef\psi="7120
\mathchardef\omega="7121
\mathchardef\varepsilon="7122
\mathchardef\vartheta="7123
\mathchardef\varpi="7124
\mathchardef\varrho="7125
\mathchardef\varsigma="7126
\mathchardef\varphi="7127

\begin{document}
 
%Title page
\pagestyle{empty}

\vskip 0.8cm
\title{\bf Quantum Mechanics without Waves: a Generalization of Classical Statistical Mechanics}
\vskip 2.0cm
\author{Marcello Cini\\ 
Dipartimento di Fisica - Universit\`a La Sapienza - Roma\\
INFM Sezione di Roma - La Sapienza}

\maketitle
\begin{abstract}
We generalize classical statistical mechanics to describe the
kynematics and the dynamics of systems whose variables are constrained by a single quantum
postulate (discreteness of the spectrum of values of at least one variable of the theory). This is
possible provided we adopt Feynman's suggestion of dropping the assumption that the probability for
an event must always be a positive number. This approach has the advantage of allowing a
reformulation of quantum theory in phase space without introducing the unphysical concept of
probability amplitudes, together with all the problems concerning their ambiguous properties.
\end{abstract}

\vskip 4truecm
\noindent
$^{(1)}$ Postal address: Piazza A.Moro, 2, 00185 Roma, Italy. E-mail: marcello.cini@roma1.infn.it.

\newpage
\pagestyle{plain}
\setcounter{page}{1}
\baselineskip=24pt

\section{Introduction}

After seventy years of Quantum Mechanics we have learned to live with complex probability
amplitudes without worrying about their lack of any reasonable physical meaning. One should not
ignore, however, that the ``wavelike'' properties of quantum objects still raise conceptual problems
on whose solutions a general consensus is far from having been reached$^{(1)(2)}$.

A possible way out of this difficulty has been implicitly suggested by Feynman$^{(3)}$, who has
shown that, by dropping the assumption that the probability for an event must always be a
nonnegative number, one can avoid the use of probability amplitudes in quantum mechanics. This
proposal, which goes back to the work of Wigner$^{(4)}$ who first introduced non positive
pseudoprobabilities to represent Quantum Mechanics in phase space, does not, however, {\it
eliminate} ``waves'',  because its starting point is the conventional mathematical framework of
Quantum Mechanics.
	
We try instead to reformulate quantum mechanics by {\it eliminating from the beginning} the concept
of probability ``waves''. This program is carried on by generalizing the formalism of classical
statistical mechanics in phase space with the introduction of a single quantum postulate
(discreteness of the spectrum of values of at least one variable of the theory), which introduces
mathematical constraints on the variables in terms of which any physical quantity can be expressed
(characteristic variables). These constraints, however, cannot be fulfilled by ordinary random
c-numbers, but are satisfied by q-numbers. The introduction of q-numbers in quantum theory is
therefore not assumed as a postulate from the beginning, but is a consequence of a well defined
physical requirement. The equations derived from these constraints allow the determination of the
expectation value of the characteristic variables for any given dispersion-free ensemble together
with the value of the physical quantity which defines it. This leads to the identification of the
characteristic variables with the Weyl operators of standard Quantum Mechanics. The whole structure
of Quantum Mechanics in phase space, including the identification of the Wigner function as the
pseudoprobability density of any quantum state, derived by Moyal in his pioneer work of
1949$^{(5)}$, is therefore deduced from a single quantum postulate without ever introducing wave
functions or probability amplitudes.  
	
\section{Classical statistical mechanics in phase space}

Consider a classical statistical ensemble of systems whose state may be defined by the values of a
couple of conjugated variables, {\bf q}, {\bf p}, which can take the values $q$, $p$, respectively.
The standard form of the joint probability density is\footnote{The constant $\hbar$ is introduced
here as a unit of action for dimensional reasons. The classical results are independent of its
value. Its identification with Planck constant will result from the comparison of quantum theory
with experiment.}

$$P(q,p) = {\rm <}\delta(q-{\bf q}) \delta(p-{\bf p}){\rm >} = (2\pi\hbar)^{-2} \int\int
dy~dk~e^{(-i/\hbar)(kq+yp)} {\rm <}e^{(i/\hbar)(k{\bf q}+ y{\bf p})}{\rm >}\eqno(1)$$
where ${\rm <}.{\rm >}$ represents the ensemble average. Similarly, any physical quantity
$A({\bf q,p})$ (for short {\bf A}) can be expressed in terms of the same variables
$e^{(i/\hbar)(k{\bf q}+y{\bf p})}$ (hereafter indicated as {\it characteristic variables}), as

$${\bf A} = \int\int dy~dk~a(k,y) e^{(i/\hbar)(k{\bf q}+ y{\bf p})}\eqno(2)$$

It is useful for our later generalization to introduce the notation ${\bf C}(k,y)$ for the
characteristic variables $e^{(i/\hbar)(k{\bf q}+y{\bf p})}$. Let us consider an ensemble in which
all the systems have the same value $\alpha$ of the physical quantity {\bf A}. Then it must be

$${\rm <} {\bf A}^2{\rm >}_\alpha = \alpha^2\eqno(3)$$
where
$$\alpha = {\rm <} {\bf A}{\rm >}_\alpha = \int\int dq ~dp ~A(q,p) ~P_\alpha(q,p).\eqno(4)$$

In order to satisfy eqs. (3) the function ${\rm <}{\bf C}(k,y){\rm >}_\alpha$ of $k$,$y$ (indicated
in the following as characteristic function of the ensemble) given by the Fourier inversion of eq.
(1) must obey the relation

$$\alpha {\rm <}{\bf C}(k,y){\rm >}_\alpha = \int\int dx~ dh~a(h-k, x-y) {\rm <}{\bf C}(h, x){\rm
>}_\alpha\eqno(5)$$
for any given value $\alpha$ of {\bf A}. To derive (5) it is crucial to use the property

$${\bf C}(k,y) {\bf C}(k',y') = {\bf C}(k+k',y+y')\eqno(6)$$

Eq. (5) is an homogeneous integral equation for the determination of the eigenvalues $\alpha$ of
{\bf A} and the corresponding eigenfunctions ${\rm <}{\bf C}(k,y){\rm >}_\alpha$. Its solutions can
be immediately obtained from its inverse Fourier transform. In terms of $A(q,p)$ (the inverse
Fourier transform of $a(k,y)$) and of $P_\alpha(q,p)$ eq. (5) shows to be no longer an integral
equation but a simple algebraic equation:

$$A(q,p) P_\alpha(q,p) = \alpha~ P_\alpha(q,p)\eqno(7)$$
which has the solutions 

$$\alpha = A(q,p)\eqno(8)$$

$$P_\alpha(q,p) = f_\alpha(\Pi(q,p)) ~~ \delta(A(q,p) - \alpha)\eqno(9)$$
with $f(\bf\Pi)$ an arbitrary function of the variable $\bf\Pi$ conjugated to {\bf A}. In fact any
other dependence of $f_\alpha$  on q and p could be expressed as a dependence on $A(q,p)$ which
would be eliminated by replacing $A(q,p)$ with $\alpha$. This arbitrariness reflects the fact that
there may be an infinity of classical ensembles in which the variable {\bf A} has the value
$\alpha$. Eq. (7) implies that, {\it given a couple of values} q,p {\it of the variables} {\bf
q,p}, {\it the variable} {\bf A} {\it has necessarily the value} (8). This seems a trivial
statement but it will turn out to be essential later.

{\it It should be noted that (9) holds for any dynamical variable, function of} {\bf q,p}. The case
of the energy is no exception, in spite of the questionable role of the time as its conjugate
variable, since the Hamiltonian {\bf H} = H({\bf q,p}) can be expressed, by means of a suitable
canonical transformation {\bf Q} =Q({\bf q,p}), {\bf P}=P({\bf q,p}), as a function E({\bf P}) of
the new momentum not depending on the new coordinate {\bf Q}. Therefore a given value P of {\bf P} 
yields a uniquely determined value E(P) of the energy. We can take therefore $\bf\Pi = {\bf Q}$. For
closed systems {\bf P} is the action variable {\bf J} = J({\bf q,p}), and E(J) is independent of the
conjugated angle variable $\bf\Theta$ = ${\bf Q}$ of ${\bf J}$.

The limiting case $f$=constant is the most interesting for the generalization we have in mind,
because {\it in this particular classical ensemble the variable ${\bf\Pi}$ is completely
undetermined}. 

In this case {\it and only in this case} the ensemble acquires a very important property. In
fact, by indicating with $\{.,.\}_{PB}$ the Poisson Bracket of {\bf A} with an arbitrary variable {\bf
B}, one has

$$\eqalign{{\rm <}\{{\bf A,B}\}_{PB}{\rm >}_\alpha &= -\int\int dp~ dq
~P_\alpha (p,q) [(\partial A/\partial q) (\partial B/\partial p) - (\partial A/\partial
p)(\partial B/\partial q)] =\cr
&= -\int\int dp~ dq
~B (p,q) [(\partial A/\partial q) (\partial P_\alpha/\partial p) - (\partial A/\partial
p)(\partial P_\alpha/\partial q)] \cr}\eqno(10)$$
namely

$${\rm <}\{{\bf A,B}\}_{PB}{\rm >}_\alpha = 0\eqno(11)$$
because when $P_\alpha$ depends only on A we have $(\partial P_\alpha/\partial \Pi) = 0$. 

Eq.(11) implies that both eqs.(3) and (11) are invariant under arbitrary infinitesimal canonical
transformations

$${\bf A}' = {\bf A} + \varepsilon \{{\bf A,B}\}_{PB}\eqno(12)$$

From eq.(11) it follows therefore that, {\it for the dispersion free ensemble in which ${\bf A}$
has the value} $\alpha$ {\it and} $\bf\Pi$ {\it is completely undetermined}, the characteristic
function satisfies, in addition to (5), also the equation 

$$\int\int dx ~dh ~a(h-k, x-y) (1/\hbar^2)(kx-hy) {\rm <}{\bf C}(h,x){\rm >}_\alpha = 0.\eqno(13)$$

Eq.(13) represents a ``classical uncertainty principle'' expressing the condition to be fulfilled
by classical ensembles having the property that when a given variable {\bf A} has the value
$\alpha$ the conjugate variable $\bf\Pi$ is undetermined. {\it Conversely, if we impose that the
characteristic function of an ensemble satisfies eqs.}(5) {\it and} (13) {\it we select only the
ensembles in which the ``uncertainty principle'' is satisfied}. 

\section{Quantum postulate}

Our reformulation of quantum theory will be based on the assumption that  eqs. (3) and (11) should
hold for any variable {\bf A}. {\it This will impose automatically for all the possible ensembles
the validity of the uncertainty principle}. However the explicit form of these equations in terms of
${\rm <}{\bf C}(h,x){\rm >}_\alpha$ given by (5) and (13) will have to be modified, because
eqs.(8) and (9) are no longer valid in quantum theory.

In {\it addition} to this first assumption, therefore, we will impose the fulfilment of an extra
postulate, based on the convinction that, instead of postulating the conventional representation of
physical quantities by means of operators in Hilbert space, it is more satisfactory to assume as a
founding stone of quantum theory the experimental fact that physical quantities exist (e.g. angular
momentum) whose possible values form a discrete set, invariant under canonical transformations,
characteristic of each variable in question.

An equally compelling physical starting point for the adoption of this postulate might be the
stability of matter. In fact this requirement implies the necessary existence of a minimum value
$E_0$ below which no lower value can be assumed by the energy of an electron-nucleus bound state.

In any case we need only assume (Quantum Postulate) that at least one variable {\bf L} exists which
has finite gaps in the continuous range L$(q,p)$ implied by its functional dependence L({\bf q,p})
on {\bf q} and {\bf p} (which can both assume any value in the continuous range $-\infty$,
$+\infty$) in which it cannot assume values except for one or more discrete values $\lambda_i$.
{\it This in fact means that, since  {\bf L} cannot have values in the range between $\lambda_i$
and $\lambda_i -
\varepsilon$, and/or between $\lambda_i$ and} $\lambda_{i}+\eta$, (with $\varepsilon, \eta$, 
finite) {\it eqs.(8) (9) do not hold in these ranges}.

As a consequence we conclude that L({\bf q,p}) cannot be expressed in the form (2), namely that the
quantum characteristic variables {\bf C}(k,y) cannot satisfy the crucial property (6).

Therefore, since by definition all variables should be expressed in terms of a unique set of
characteristic variables, we conclude that for {\it all} the variables\footnote{We change the
notation from ${\bf A}$ to ${\bf\hat{A}}$ in order to distinguish the variables satisfying
the Quantum Postulate from those satisfying (2)} ${\bf\hat{A}}$, eq.(2) should be replaced
by

$${\bf\hat{A}} = \int\int dy ~dk~ a(k,y) {\bf\hat{C}}(k,y)\eqno(14)$$
with a set of characteristic variables ${\bf\hat {C}}(k,y)$ obeying a new rule of multiplication
replacing eq.(6).

In order to find the required modification of eq.(6) we start by asking how the eigenvalue equation
(5) should be modified in order to allow, besides (instead or in addition to) a continuous range of
possible values, also for the existence of discrete  eigenvalues $\lambda_i$ of $\bf\hat{L}$.
This amounts to say that its Fourier transform should no longer reduce to the algebraic relation
(7) but should become a true Fredholm homogeneous integral equation, which, as is well known, has
the property, under suitable conditions, of allowing for the existence of discrete eigenvalues.

Whether a given variable ${\bf\hat{A}}$ will actually have eigenvalues belonging to a discrete
or a continuous (or even both) spectrum will depend on its functional dependence on ${\bf\hat{p}}$
 and ${\bf\hat{q}}$. There are in any case some stringent requirements that the modified kernel
should satisfy to attain this goal, namely:

\noindent
a) the basic information on the functional dependence of ${\bf\hat{A}}$ on ${\bf\hat{p}}$ and
${\bf\hat{q}}$ contained in the kernel a(h-k, x-y) should remain unaltered;

\noindent
b) the correlation between the couple of variables $h,x$ and $k,y$ which is necessary in order to
transform eq.(5) into a true Fredholm integral equation should be universal, namely {\it
independent of the variable chosen and of the state considered};

\noindent
c) the classical kernel should be recovered when $k = y$ =0 because eq. (5) for ${\rm
<}{\bf\hat{C}}(0,0){\rm >}_\alpha = 1$ should give eq.(4) which must still be valid;

\noindent
d) the classical kernel should be recovered also for $h = x$ =0 because the relation (6) should
still be valid when $k = hÕ$ and $x = yÕ$.

The simplest (and from this point of view unique) way to satisfy a) and b) is to multiply the
classical kernel $a(h-k, x-y)$ by a factor g$(kx-hy)$ whose argument is unambiguously fixed by the
requirement that, for dimensional reasons, $x$ should be correlated to $k$ and $h$ to $y$.
Furthermore in order that c) and d) are fulfilled, it must be g(0)=1. The modified integral
equation replacing eq. (5) should therefore read
$$\alpha~_i {\rm <}{\bf\hat{C}}(k,y){\rm >}_i = \int\int dx~ dh ~a(h-k, x-y) ~{\rm g}(kx-hy) {\rm
<}{\bf\hat{C}}(h, x){\rm >}_i\eqno(15)$$
 
Eq. (15) has a first important consequence. In fact the condition (3), which may be rewritten in
terms of the new variables ${\bf\hat{C}}(k,y)$ in the form
		
$$\eqalign{&\int\int dydk ~a(k,y) \int\int dy'dk' ~a(k',y'){\rm <}{\bf\hat{C}}(k,y)
{\bf\hat{C}}(k',y'){\rm >}_i\cr
& = \alpha_i\int\int dydk ~a(k,y) {\rm <}{\bf\hat{C}}(k,y){\rm >}_i\cr}\eqno(16)$$
leads to eq. (15) only if eq. (6) is replaced by
$$(1/2)[{\bf\hat{C}}(k,y) {\bf\hat{C}}(k',y') + {\bf\hat{C}}(k',y') {\bf\hat{C}}(k,y)] = {\rm
g}(ky'- k'y) {\bf\hat{C}}[(k+k'), (y+y')]\eqno(17)$$

{\it This equation, however, cannot be satisfied by ordinary c-numbers. This means that, if we want
to allow for the existence of discrete values of at least one variable ${\bf\hat{L}}$ we are
forced to represent \underbar{all} the variables ${\bf\hat{A}}$ by means of q-numbers}. We need
not, however assume for these q-number variables other properties except that they exist and that
(17) is satisfied. {\it This means that the mathematical nature of the entities needed to represent
the quantum variables is a consequence of the physical property represented by our Quantum
Postulate, and not viceversa, as the conventional view of reality underlying the conventional
axiomatic formulation of Quantum Mecchanics assumes}.  

We need not give any new rule in order to define the symbol ${\rm <}{\bf\hat{C}}(k,x){\rm
>}_i$ in terms of operators and state vectors, because on the one hand its physical meaning is {\it
by definition} the same of its classical counterpart ${\rm <}e^{(i/\hbar )(k{\bf q}+y{\bf p})}{\rm
>}_i$ namely the mean value of the characteristic variable in the ensemble in which the variable
${\bf\hat{A}}$ has the value $\alpha_i$, and on the other hand its explicit expression will be
{\it derived} by solving eq. (15) together with the analogous quantum generalization of eq. (13)
which we will now proceed to write down.

In order to fulfill the condition that the eigenvalues $\alpha_i$ of ${\bf\hat{A}}$ should be
invariant under the canonical transformations (12) one must in fact impose that eq. (11) should
hold. However, if we use for the new characteristic variables ${\bf\hat{C}}(k,y)$ the
classical PBÕs of the old variables
$$\{e^{(i/\hbar )(k{\bf q}+y{\bf p})}, e^{(i/\hbar )(k'{\bf q}+y'{\bf p})}\}_{PB} =
[(k'y-ky')/\hbar^2] e^{(i/\hbar )[(k+k'){\bf q}+(y+y'){\bf p}]}\eqno(18)$$
we immediately see that eq. (17) is no longer invariant under (12) which has therefore to be
replaced by
$${\bf\hat{A}}' = {\bf\hat{A}} + \varepsilon \{{\bf\hat{A}},{\bf\hat{B}}\}_{QPB}\eqno(19)$$

We have therefore to {\it derive} the corresponding quantum Poisson Brackets (QPB) of the two
variables ${\bf\hat{C}}(k,y)$, ${\bf\hat{C}}(k',y')$ from the condition of invariance of (17)
under (19). Here again we need not introduce explicitly the standard definition of the PB's of these
q-numbers in terms of operators. On the contrary, their form will be {\it obtained} as a
consequence of our formalism. We will only need to define  QPB's, for consistency with eq. (17), by
means of the following generalization of the classical PBÕs
$$\{{\bf\hat{C}}(k,y), {\bf\hat{C}}(k',y')\}_{QPB} = {\rm f}(ky'-k'y) {\bf\hat{C}}[(k+k'),
(y+y')]\eqno(20)$$ where f$(\lambda)$ is an odd function of its argument satisfying, for
consistency with eq. (18), the condition $\lim_{\lambda\to 0} {\rm f}(\lambda)  = -\lambda/\hbar^2$ 

From (17) and (20) we now obtain immediately
$$\int\int dx ~dh~ a(h-k, x-y) ~{\rm f}(kx-hy) {\rm <}{\bf\hat{C}}(h,x){\rm >}_i =
0\eqno(21)$$

This is the required generalization of eq. (13). 

The further step required to complete our formalism is the determination of the functions f(.) and
g(.). The knowledge of these functions will then allow the explicit derivation of the characteristic
function ${\rm <}{\bf\hat{C}}(k,y){\rm >}_i$ and the corresponding eigenvalue $\alpha_i$ of
${\bf\hat{A}}$ for any state ${\rm< >_i}$ by solving eqs. (15) and (21). This goal is easily
attained by imposing the condition that both relations (16) and (20) should be invariant under the
canonical transformations (19). This condition leads in fact to the two equations 
$${\rm f}(\lambda) {\rm f}(\mu -\nu) + {\rm f}(\mu) {\rm f}(\nu -\lambda) + {\rm f}(\nu) {\rm
f}(\lambda -\mu) =0\qquad {\rm (Jacobi~~ identity)}\eqno(22)$$

$${\rm g}(\lambda) {\rm f}(\mu +\nu) = {\rm g}(\lambda +\mu) {\rm f}(\nu) + {\rm g}(\lambda -\nu)
{\rm f}(\mu)\eqno(23)$$

These equations have the following solutions 
$${\rm g}(ky'-k'y) = \cos [b(ky'-k'y)/\hbar]~;\quad {\rm f}(ky'-k'y) = (1/b\hbar)
\sin[b(k'y-ky')/\hbar]\eqno(24)$$
with $b$ a parameter which is still undetermined. It should be stressed that the classical
statistical theory is {\it not} recuperated by making $\hbar \to 0$, but by letting the adimensional
parameter $b \to 0$ (absence of correlations). However, although $b \to 0$ is a valid mathematical limit
for the expressions (25), $b=0$ and $b\not= 0$ yield two radically different theories, because in
the first case the variables are c-numbers while in the second one they are q-numbers.

The solution ${\rm <}{\bf\hat{C}}(k,y){\rm >}_i$ of eqs. (15) (21) will now yield easily the
corresponding expression for $P_i(q,p)$ by means of
$$P_i(q,p) = (2\pi\hbar)^{-2}\int\int~ dy~ dk~ e^{(-i/\hbar)(kq+yp)} {\rm <}{\bf C}(k,y){\rm
>}_i\eqno(25)$$
	
Before discussing the properties of this (pseudo)probability density we will however work out the
results of our theory in some simple cases.

\section{Simple examples}

We will first solve the two equations (15) (21) for the variables ${\bf\hat{q},\bf\hat{p}}$ and
successively for the energy ${\bf\hat{H}} = (1/2) {\bf\hat{p}}^2 + (1/2) \omega^2
{\bf\hat{q}}^2$ of the harmonic oscillator. This will show explicitly how the formalism leads
both to the existence of variables whose eigenvalues belong to a continuous range as well as of
other ones with a discrete spectrum.

\noindent
1. {\it Variable} ${\bf\hat{q}}$. From the classical expression (2) one obtains
$$a_q(k,y) = \int\int dq~ dp~ q \exp[-i(py+qk)/\hbar] = i\hbar \delta(y) [\partial\delta
(k)/\partial k]\eqno(26)$$

The eigenvalue equation (15) reads
$$\eqalign{q_o{\rm <}{\bf\hat{C}}(k,y){\rm >}_{q_o} &= i\hbar\int\int dx~dh~\delta(x-y)
[\partial\delta(h-k)/\partial h] {\rm g}(kx-hy) {\rm <}{\bf\hat {C}}(k,y){\rm >}_{q_o} =\cr
&= -i\hbar [\partial {\rm <}{\bf\hat{C}}(k,y){\rm >}_{q_o} /\partial k]\cr}\eqno(27)$$
because ${\rm g}(0) = 1$ and $[\partial {\rm g}(\lambda)/\partial\lambda]_{\lambda =0} = 0$, where
$q_0$ is the value of ${\bf\hat{q}}$ which labels the state. The solution of (27) is
$${\rm <}{\bf\hat{C}}(k,y){\rm >}_{q_o} = \exp[ik q_o /\hbar] \phi(y)\eqno(28)$$
with $\phi(y)$ an arbitrary function. On the other hand eq. (21) reads
$$0 =  i\hbar\int\int dx ~dh~ \delta(x-y) [\partial\delta(h-k)/\partial h] f(kx-hy) {\rm
<}{\bf\hat{C}}(h,x){\rm >}_{q_o} = y{\rm <}{\bf\hat{C}}(k,y){\rm >}_{q_o}\eqno(29)$$ 
because ${\rm f}(0) = 0$ and $[\partial {\rm f}(\lambda)/\partial\lambda]_{\lambda =0} =
1/\hbar^2$. Eq. (29) gives immediately
$$\phi(y) = \delta(y)\eqno(30)$$
By introducing (28) (30) in eq. (25) one obtains 
$$P_{q_o} (q,p) = (2\pi\hbar)^{-1} \delta(q-q_o)\eqno(31)$$
which coincides with the classical probability density of the ensemble in which the variable
${\bf\hat{q}}$ has the value $q_o$.

This shows that the possible values of the quantum variable ${\bf\hat{q}}$ span the same
continuous range from $-\infty$ to $+\infty$ of the classical variable ${\bf\hat{q}}$. This is
because the solution of (27) and (29) involves only the classical limits of g(.) and f(.) and does
not depend on the actual value of  $b$. In this case the QPB's coincide with the classical PB's.
\vskip 6pt

\noindent
2. {\it Variable} ${\bf\hat{p}}$. The complete symmetry between ${\bf\hat{q}}$ and ${\bf\hat{q}}$
allows us to write 
$${\rm <}{\bf\hat{C}}(k,y){\rm >}_{p_o} = \exp[iyp_o /\hbar] \delta(k)\eqno(32)$$
namely
$$P_{p_o} (q,p) = (2\pi\hbar)^{-1} \delta(p-p_o)\eqno(33)$$

\noindent
3. {\it Variable} ${\bf\hat{H}} = (1/2){\bf\hat{p}}^2 + (1/2)\omega^2 {\bf\hat{q}}^2$. From
the classical expression we obtain
$$\eqalign{h(k,y) &= \int\int dq ~dp (1/2)[p^2 + \omega^2 q^2] \exp[-i(py+qk)/\hbar] =\cr
&= -(\hbar^2/2) [\delta(k) \partial^2\delta(y)/\partial y^2 + \omega^2 \delta(y)
\partial^2 \delta(k)/\partial k^2]\cr}\eqno(34)$$

Eq. (15) reads
$$\eqalign{&E_o {\rm <}{\bf\hat{C}}(k,y){\rm >}_{E_o} =\cr
&= -(\hbar^2/2)\int\int dx~ dh~ \delta(x-y) \delta(h-k)[\partial^2/\partial x^2 + \omega^2
\partial^2/\partial h^2] [{\rm g}(kx-hy) {\rm <}{\bf\hat{C}}(h,x){\rm >}_{E_o}]\cr}\eqno(35)$$
Since, from (24) we have 
$$\eqalign{&[\partial^2 {\rm g}(hy-kx)/\partial x^2]_{x=y; h=k} = -k^2 b^2/\hbar^2\cr 
&[\partial^2 {\rm g}(hy-kx)/\partial h^2]_{x=y; h=k} = -y^2 b^2/\hbar^2\cr}\eqno(36)$$
we obtain
$$E_o {\rm <}{\bf\hat{C}}(k,y){\rm >}_{E_o} = (1/2)[k^2 b^2 - \hbar^2
\omega^2\partial^2/\partial k^2 +
\omega^2 y^2 b^2 - \hbar^2 \partial^2/\partial k^2] {\rm <}{\bf\hat{C}}(k,y){\rm
>}_{E_o}\eqno(37)$$

From eq. (21) we obtain
$$[k\partial /\partial y - \omega^2 y\partial /\partial k]{\rm <}{\bf\hat{C}}(k,y){\rm
>}_{E_o} = 0\eqno(38)$$

Eq. (38) can be solved by setting
$${\rm <}{\bf\hat{C}}(k,y){\rm >}_{E_o} = F(k) G(y)\eqno(39)$$
leading to
$$F(k) = \exp[\mu k^2/\omega^2]\qquad\qquad G(y) = \exp[\mu y^2]\eqno(40)$$

Introducing (39) and (40) in (37) we easily find (since ${\rm <}{\bf\hat{C}}(0,0){\rm >} = 1)$
$$E_o =  b\hbar~\omega\eqno(41)$$

$${\rm <}{\bf\hat{C}}(k,y){\rm >}_{E_o} = \exp[- bk^2/2\hbar\omega] \exp[-by^2\omega/2\hbar]
\eqno(42)$$

By introducing (42) in (25) we obtain
$$P_{E_o} (q,p) = (1/2\pi\hbar b) \exp[-p^2/2\hbar\omega b] \exp[-q^2 \omega/2\hbar b]\eqno(43)$$

For the excited states the separability condition (39) does not hold. Eqs. (37) andÊ(38) are
however sufficient to determine completely the corresponding characteristic functions and
eigenvalues$^{(10)}$. 
	
\section{The uncertainty principle}

We will finally discuss the properties of the (pseudo)probability densities $P_i(q,p)$ given by
(25). This will also allow us to determine the parameter $b$.

We start by writing eq. (21) for both ${\rm <}{\bf\hat{C}}(k,y){\rm >}_i$ and ${\rm
<}{\bf\hat{C}} (k,y){\rm >}_j$ for $i\not= j$; we multiply the first one by ${\rm
<}{\bf\hat{C}}(-k,-y){\rm >}_j$ and the second one by ${\rm <}{\bf\hat{C}}(-k,-y){\rm >}_i$
and finally integrate over $k, y$. By subtracting the second equation from the first one we obtain
$$0 = (\alpha_i - \alpha_j) \int\int dy ~dk ~{\rm <}{\bf\hat{C}}(-k,-y){\rm >}_i~{\rm
<}{\bf\hat{C}} (k,y){\rm >}_j\eqno(44)$$

This amounts to writing\footnote{We assume that the spectrum is nondegenerate.} 
$$\int\int dy ~dk {\rm <}{\bf\hat{C}}(-k,-y){\rm >}_i~{\rm <}{\bf\hat{C}}(k,y){\rm >}_j = N
\delta_{ij}\eqno(45)$$
where $N$ is a normalization constant, having the dimensions of an action, independent of the
variable ${\bf\hat {A}}$ and of the state ${\rm <>}_i$. From (25) and (45) we obtain
$$\int\int dq~ dp [(2\pi\hbar)^2 P_i(q, p) /N] P_j(q, p) = \delta_{ij}\eqno(46)$$

{\it At this stage we have to fix our unit of action} $2\pi\hbar$. To this purpose we compare (46)
with its semiclassical limit given by the old theory of quanta of Planck and Bohr where the volume
of the region of phase space in which the classical $A(q,p)$ has the value $\alpha_i$ and all the
points $q,p$ have equal constant probability $K_i$ inside it and zero probability outside, is
assumed to be equal to Planck's constant $(2\pi\hbar)$. Then in this semiclassical theory, we have,
for the normalization of probability
$$2\pi\hbar~ K_i =1\eqno(48)$$
and for (46)
$$(2\pi\hbar)^3 ~K_i^2 /N = 1\eqno(49)$$
from which we get $N = 2\pi\hbar$. Eq. (46) becomes therefore 
$$\int\int dq~ dp ~P_i(q, p) P_i(q, p) = (2\pi\hbar)^{-1} \equiv P_{av}\eqno(50)$$

The last step of our work is now the determination of the parameter $b$. In fact it is immediate to
see that, introducing into eq. (50) the expression (43) for $P_{E_o}~ (q,p)$ of the harmonic
oscillator ground state, one obtains $b=1/2$. Since this value is independent of the variable and of
the state chosen, this result is wholly general and consequently our reformulation of quantum
theory is complete.

Eq. (50) expresses a new form of the uncertainty principle for position and momentum. In fact, by
introducing in the normalization condition the mean value $P_{av}$ defined by this equation, we
obtain
$$\int\int dq ~dp~  P(q,p) = P_{av} \delta q~\delta p = 1\eqno(51)$$
where $\delta q~\delta p$ is the volume of phase space in which $P(q,p)$ is replaced by $P_{av}$ and
is zero outside. We then immediately obtain
$$\delta q~\delta p = 2\pi\hbar .\eqno(52)$$

It is important to stress that (52) does not have the form of the conventional Heisenberg
inequality, which gives no upper limit to the possible value of the uncertainty product $\Delta
q~ \Delta p$ of the mean square values of $q$ and $p$, but involves only its minimum value. We will
return on the implications of this difference in the discussion.
		
\section{The dynamical evolution}

We finally indicate how the dynamical evolution of the pseudoprobability distribution $P(q,p)$ in
any given state given by (25) can be worked out. It is sufficient to use the Hamiltonian
${\bf\hat {H}}$
$${\bf\hat{H}} = \int\int dy~ dk~ h(k,y) {\bf\hat{C}}(k,y)\eqno(53)$$
as the generator of the infinitesimal displacement in time
$$d{\bf\hat{C}}(k,y)/dt = \{{\bf\hat{C}}(k,y), {\bf\hat{H}}\}_{QPB} = \int\int dx~ dj~
h(j-k,z-y) ~f(jy-kx) ~{\bf\hat{C}}(j,x)\eqno(54)$$
Eq. (54) yields a Chapman-Kolmogorov equation for the time dependence of the pseudoprobability
density
$P(q,p;t)$:
$$(d/dt) P(q,p;t) = \int\int dq'dp' K(q,p; q',p') P(q',p';t)\eqno(55)$$
In the classical limit eq.(55) reduces to the Liouville equation.	
 
{\it We have therefore attained our goal, namely the construction of a formal probabilistic theory
(with the generalization of probabilities to negative values according to Feynman's interpretation)
of the quantum world  in phase space by means of a straightforward generalization of classical
statistical mechanics.}
	
\section{Comparison with the conventional formulation of Quantum Mechanics}

The present formulation of quantum theory is clearly identical to the conventional formalism of
Quantum Mechanics. In fact if we consider the Weyl operator
$${\underline{\rm{\bf C}}}(k,y) =
e^{i({\underline{\rm{\bf p}}}y+k{\underline{\rm{\bf q}}})/\hbar}\eqno(56)$$  
where ${\underline{\rm{\bf p}}}$  and  ${\underline{\rm{\bf q}}}$ are the momentum and position
operators satisfying the usual commutation relation 
$[{\underline{\rm{\bf q}}},{\underline{\rm{\bf p}}}] = i\hbar$, one
finds immediately that
$${\bf\hat{C}}(k,y) = {\underline{\rm{\bf C}}}(k,y)\eqno(57)$$ 
Therefore, if $\vert\psi {\rm >}$ is the state vector corresponding to our state ${\rm <. >}$ we
have
$${\rm <}{\bf\hat{C}}(k,y) {\bf\hat{C}}(k',y') + {\bf\hat{C}}(k',y') {\bf\hat{C}}(k,y){\rm >}
= 2Re{\rm <}
\psi\vert {\underline{\rm{\bf C}}}(k,y) {\underline{\rm{\bf C}}}(k',y')\vert\psi {\rm >}\eqno(58)$$
$${\rm <}\{{\bf\hat{C}}(k,y), {\bf\hat{C}}(k',y')\}_{QPB}{\rm >} = (2/\hbar) {\rm Im <}
\psi\vert {\underline{\rm{\bf C}}}(k,y) {\underline{\rm{\bf C}}}(k',y')\vert\psi{\rm >}\eqno(59)$$
From (57) (58) it follows also that
$$P(q,p) = W(q,p)\eqno(60)$$
where $W(q,p)$ is the Wigner function$^{(4)}$ of the state $\vert\psi{\rm >}$. Eq. (55) coincides
therefore with the standard equation for the time evolution of the Wigner function. This result
shows that this function has a privileged status among other functions $^{(6)}$ used in the
literature to describe Quantum Mechanics in phase space, because it can be derived directly from
our quantum postulate.

\section{Discussion}

The physical meaning of negative probabilities is well clarified by Feynman's own words: {\it ``It
is that a situation for which a negative probability is calculated is impossible, not in the sense
that the chance for its happening is zero, but rather in the sense that the assumed conditions of
preparation or verification are experimentally unattainable.''} Admittedly, as he recognizes, a
``strong mental block'' against this extention of the probability concept is widespread. Once this
has been overcome, however, the present formulation of quantum theory has several advantages.

First of all, as already anticipated in the introduction, many paradoxes typical of the
wave-particle duality disappear. On the one hand in fact, as already shown by Feynman, it becomes
possible to express the correlations between two distant particles in terms of the product of two
probabilities independent from each other$^{(3)(7)}$. All the speculations on the nature of an
hypothetical superluminal signal between them becomes therefore meaningless. On the other hand the
long time debated question about the meaning of the superposition of state vectors for macroscopic
objects (Schr\"odinger's cats) may also be set aside as equally baseless, together with the many
proposals of detection of ``empty waves''. It is not the practical use of the formalism of Quantum
Mecanics, of course, which is put in question. However, from a conceptual point of view, the
elimination of the waves from quantum theory is in line with the procedure inaugurated by Einstein
with the elimination of the aether in the theory of electromagnetism.  

Secondly, this approach eliminates the conventional hybrid procedure of describing the dynamical
evolution of a system, which consists of a first stage in which the theory provides a deterministic
evolution of the wave function, followed by a hand made construction of the physically meaningful
probability distributions. If the probabilistic nature of the microscopic phenomena is fundamental,
and not simply due to our ignorance as in classical statistical mechanics, why should it be
impossible to describe them in probabilistic terms from the very beginning? 

The third advantage is connected with the possibility of dissipating the ambiguity of the
conventional theory about two physically different aspects of the quantum uncertainties inherent to
the Heisenberg inequality. It has been recognized in fact that this inequality contains two
contributions of different origin$^{(8)}$. Its minimum value is in fact an {\it ontological} 
uncertainty, of quantum nature, while the contribution exceeding this minimum is of {\it
epistemic}  nature, namely expresses a statistical effect due to imperfect knowledge of reality. In
fact, while the irreducible quantum contribution requires that a reduction of $\Delta x$ should
necessarily imply a simultaneous increase of $\Delta p$ (or viceversa), for the statistical
contribution both uncertainties can be reduced at the same time by more accurate measurements until
the minimum value is reached. In the present formulation of quantum theory, however, only the
quantum ontological uncertainties are present, without any spurious statistical contribution. This
is because the uncertainty principle in our theory is given by the equality (52), involving only
the minimum value of the Heisenberg inequality.

The last, but not least, appeal of this approach is that it may be cosidered as a conceptual
``Gestalt switch'' of the type suggested by Thomas Kuhn$^{(9)}$ concerning the status of the ``Laws
of Nature''. A switch from the ``autocratic'' rule that {\it the Laws prescribe everything that must
happen} to the ``democratic'' principle that {\it anything which is not forbidden by the Laws may
happen}. If chance has an irreducible origin the fundamental laws cannot prescribe everything: they
can only express constraints following from stability requirements of matter, or prohibitions
deriving from symmetry properties of the Universe, or general principles warranting the existence
of patterns of order. In other words they should allow for the occurrence of different events under
equal conditions. If this is true, it becomes meaningless to ask: {\it how can this event happen?} 
The answer can only be: {\it it happens because it is not forbidden}. The language of probability,
suitably adapted to take into account all the relevant constraints, seems therefore to be the only
language capable of expressing this fundamental role of chance.

\section*{Acknowledgement}

The contribution of my friend and colleague Gianni Jona-Lasinio has been crucial in clarifying the
mathematical nature of the characteristic variables and in giving to the formulation of the theory
a tight logical structure. I wish therefore to express my gratitude for his constructive criticism,
without which this paper would not have been accomplished. Illuminating discussions with Francesco
Guerra and Maurizio Serva are also gratefully acknowledged. 
\vfill\eject

\end{document}